\documentclass{article}
\pdfoutput=1
\usepackage{cite}
\usepackage{amsmath}
\usepackage{amssymb}

\usepackage{dsfont}
\usepackage{slashed}
\usepackage{multicol}
\usepackage{caption}
\usepackage[top=3cm,bottom=3.5cm,left=2.75cm,right=2.75cm]{geometry}
\usepackage{graphicx}
\usepackage{lettrine}

\newenvironment{Figure}
  {\par\medskip\noindent\minipage{\linewidth}}
  {\endminipage\par\medskip}

\usepackage{abstract}

\usepackage{titlesec}
\titleformat{\section}[block]{\large\scshape\centering}{\thesection.}{1em}{}
\titleformat{\subsection}[block]{\scshape}{\thesubsection.}{1em}{}

\title{\vspace{-15mm}%
	\fontsize{20pt}{5pt}\selectfont
	\textbf{The Problem with False Vacuum Higgs Inflation}
	}	
\author{
	\large
	\textsc{Malcolm Fairbairn\footnote{malcolm.fairbairn@kcl.ac.uk},Philipp Grothaus\footnote{philipp.grothaus@kcl.ac.uk}, and Robert Hogan\footnote{robert.hogan@kcl.ac.uk}}
	\\[2mm]
	\normalsize	Physics, Kings College London, Strand, London WC2R 2LS, UK
	\vspace{-5mm}
	}
\date{}

\begin{document}

\maketitle
\begin{abstract}
\noindent We investigate the possibility of using the only known fundamental scalar, the Higgs, as an inflaton with minimal coupling to gravity. The peculiar appearance of a plateau or a false vacuum in the renormalised effective scalar potential suggests that the Higgs might drive inflation. For the case of a false vacuum we use an additional singlet scalar field, motivated by the strong CP problem, and its coupling to the Higgs to lift the barrier allowing for a graceful exit from inflation by mimicking hybrid inflation. We find that this scenario is incompatible with current measurements of the Higgs mass and the QCD coupling constant and conclude that the Higgs can only be the inflaton in more complicated scenarios.

\end{abstract}
\section{Introduction}
\begin{multicols}{2}
A period of exponential expansion in the early Universe solves the horizon, flatness and monopole problem as well as sourcing the seeds of structure formation. The spectrum of scalar perturbations predicted from such inflationary theory has been measured many times, most recently to an impressive accuracy by the Planck satellite \cite{Planck}. 

The recently reported observation of primordial B-modes in the polarization of the CMB by the BICEP-2 experiment \cite{BICEP} may turn out to be the most convincing evidence of inflation to date.  Although the Planck data has made some steps in selecting from the various models that can produce inflation we are still a long way from pinning down what features the precise microscopic mechanism responsible for inflation would have to have.  

What is, however, common to almost all models is the presence of a scalar inflaton. The discovery of the Higgs boson, $h$, by the ATLAS \cite{ATLAS1207} and CMS \cite{CMS1207} collaborations is the first (seemingly \cite{Ellis1303}) fundamental scalar we have detected. It is therefore natural to ask whether the Higgs can play the role of the inflaton. A naive first answer would be that it cannot because it is well known that for $V(\phi)\simeq \frac{1}{4}\lambda \phi^4 $ the measured spectrum of perturbations requires\footnote{This requirement is to fit the perturbations for $N=60$ e-folds before the end of inflation. This model is also in tension with Planck's $n_S -r$ plane constraints \cite{Planck}, where $n_S$ is the spectral index and $r$ is the tensor-to-scalar ratio} the quartic coupling $\lambda \simeq 10^{-13}$ whereas the measured Higgs mass requires $\lambda \sim 0.13$. This, however, neglects the effect of quantum corrections. Properly considered, these effects can lead to substantial modifications to the tree-level potential and a significant scale dependence of $\lambda$.

For a finely chosen mass of the top quark it is possible, as shown in \cite{Isidori0712}, that the effective Higgs potential develops a flat part at large field values or even a second, local minimum, also called a false vacuum. Remarkably these features appear at approximately the correct scale to generate the observed perturbations which suggests the Higgs does indeed have a role to play in inflation.

Recently there has been a lot of interest in using the Higgs as the inflaton in the context of a non-minimal coupling to gravity~\cite{Bezrukov0710,Bezrukov0812,Bezrukov0812a,Barbon0903,Bezrukov0904,Lerner0912,Burgess1002,Bezrukov1008,Giudice1010}. It is worth noting that quantum corrections may reduce the predictiveness of such models \cite{burgess1402} and should be taken into account. Additionally, if the recent measurement by the BICEP collaboration proves to be true then these models will be put under pressure~\cite{Cook1403} (for a possible way out see the recent works~\cite{Bezrukov1403,Hamada1403} that rely on similar tunings of the Higgs potential). Here we don't consider any such coupling and so refer to it as minimal Higgs inflation.

In this paper we will investigate how the plateau or the false vacuum could be used to explain the inflationary phase of the universe. To do so we will first look at the situation where there is a plateau in the potential and see whether the Higgs can inflate the universe on its own by slowly rolling down the plateau. The case of a false vacuum in the potential demands a mechanism for a graceful exit from inflation. Therefore, we extend the model and add an additional scalar field, $s$, which can lift the Higgs out of its local minimum. The strong CP-problem motivates the existence of such an additional scalar field and it is worth investigating if such a mechanism can give successful inflation.

Our calculation improved upon a previous treatment in~\cite{masina1204} by considering the full 3-loop renormalisation group equation (RGE) improved 2-loop effective potential~\cite{degrassi1205,buttazzo1307}, including the 1-loop RGE's for the new scalar field and its threshold effect at the matching scale. Also, we account for the movement of the Higgs during inflation and further address a degeneracy in the initial depth of the false vacuum. We will see that these improvements can dramatically affect the conclusions.

The structure of the paper is as follows. In section 2 we discuss the RGE improved effective potential and attempt to use the resulting plateau for inflation. In section 3 we discuss the possibility of false vacuum inflation which is the main focus of this paper. Finally, in section 4 we present our conclusions.

\section{Plateau Inflation}
In \cite{degrassi1205,buttazzo1307} a state of the art 3-loop RGE improved 2-loop effective potential for the Standard Model Higgs was presented and discussed. This calculation showed that, within the current experimental errors on the Higgs and top masses, we appear to be living in a very special Universe. In figure 5 of \cite{degrassi1205} we can see that the current experimental data places the electroweak vacuum at the boundary between stable and meta-stable. The possibility of instability/meta-stability, as has been much discussed \cite{Isidori0104,Isidori0712,ArkaniHamed0801,Ellis0906,EliasMiro1112,degrassi1205,buttazzo1307}, is largely a result of the sizeable negative contribution of the top yukawa coupling to the beta function of the Higgs quartic coupling, $\lambda_h$, which can cause $\lambda_h$ to become negative at some high scale, creating an additional AdS vacuum into which we might tunnel. We appear to be safe from a catastrophic tunnelling event, however, because the lifetime of our vacuum is much greater than the current age of the Universe \cite{degrassi1205} (note that when the Higgs is not the inflaton, its dynamics during inflation might drastically reduce this lifetime \cite{Espinosa0710,Lebedev1210,Kobakhidze1301, Fairbairn1403}). The proximity of the current experimental data to the boundary between stable and meta-stable is a result of the peculiar fact that both $\lambda_h$ and $\beta_{\lambda_h}$ can vanish at the same scale, which is highly non-trivial and merits investigation.
\begin{Figure}
\centering
\includegraphics[width=\linewidth]{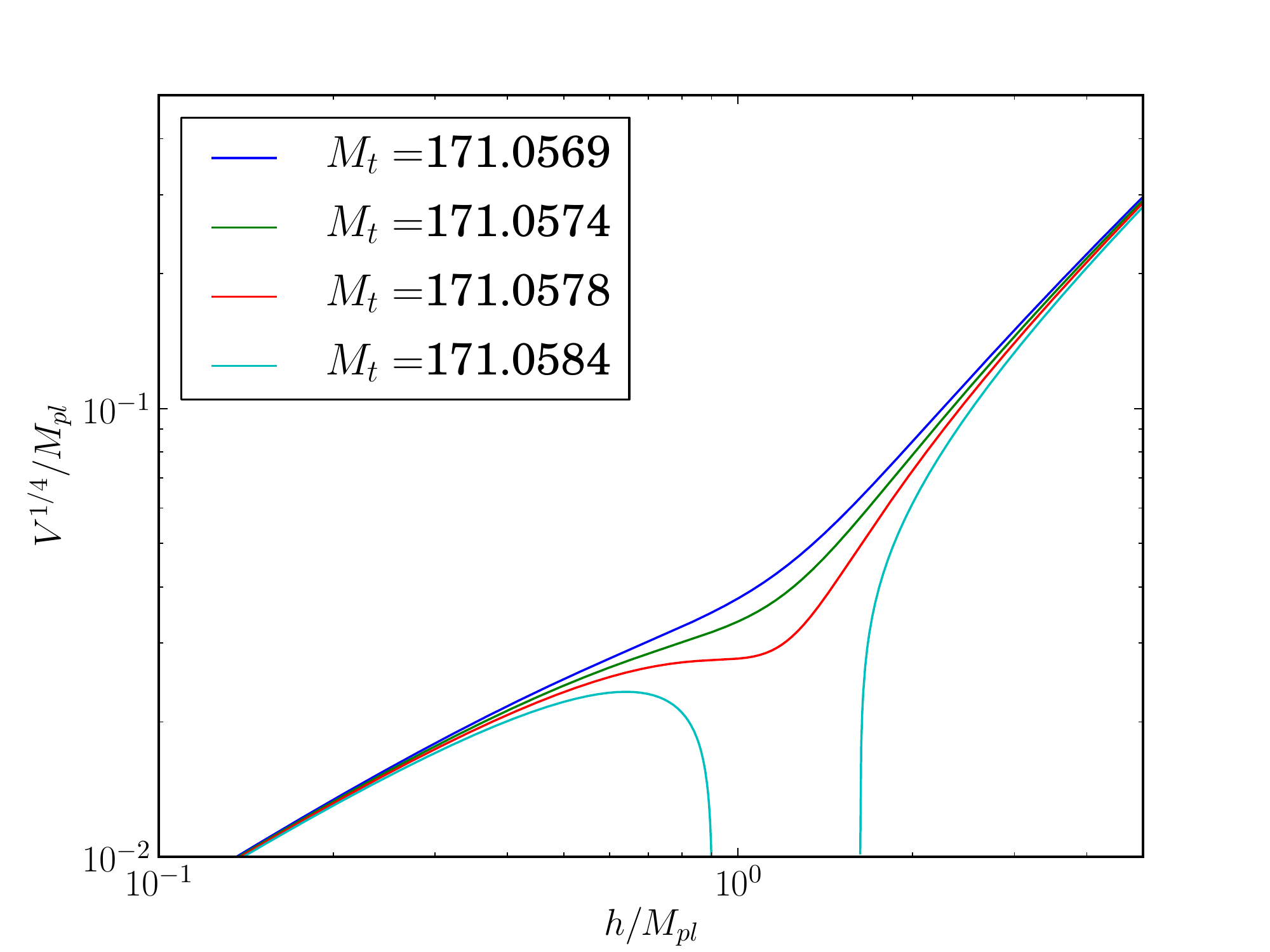}
\captionof{figure}{\it The figure shows the effective potential for $M_h=125$ GeV. The top mass is tuned in order to show the appearance of a plateau or an instability. The four curves plotted differ by $0.5$ MeV in $M_t$. Here $M_{pl} =2.345\times 10^{18}$ GeV is the reduced Planck mass.}
\label{fig: plateau}
\end{Figure}
One consequence is the development of plateau in the Higgs potential that could lead to slow roll inflation \cite{Isidori0712} and, remarkably, it appears at approximately the correct scale to generate the observed perturbations. In figure \ref{fig: plateau} we can see the effect of tuning the top mass on the effective potential. The figure shows that tuning on the order of $0.1$-$1$ MeV can interpolate between and stable and meta-stable vacuum. At the boundary of this transition we see the appearance of a plateau.
In order to test the suitability of this scenario for inflation we can start the field above the plateau and let it roll down the potential  and calculate the e-folds. The field, $h$, will evolve according to the field equations,
\begin{equation}
\ddot{h}+3H\dot{h}=\frac{dV_{\text{eff}}}{dh},
\end{equation}
with
\begin{equation}
H=\frac{1}{M_{pl}}\sqrt{\frac{\rho}{3}} \hspace{.7cm} \text{and}  \hspace{.7cm} \rho=\frac{1}{2}\dot{h}^2+V_{\text{eff}} .
\end{equation} 
Here we have 
\begin{equation}
V_{\text{eff}} =\frac{1}{4}\lambda_{\text{eff}}h^4,
\end{equation}
where $\lambda_{\text{eff}}$ contains the 3-loop RGE's and the 2-loop corrections to the effective potential such that when we choose $h$ as the renormalization scale $\lambda_{\text{eff}}$ becomes a function of $h$.
The total number of e-folds is then given by,
\begin{equation}
N_{tot}=\int_{t_{i}}^{t_{f}} H dt.
\end{equation}

The result is shown in figure \ref{fig: plateau_inflation}. We see that in order to get the required e-folds (50-60) to solve the horizon problem we need $M_h \gtrsim 129$ GeV which is inconsistent with the value observed at the LHC.
It is possible to try to relieve this constraint by, say, introducing another scalar that mixes with the Higgs such that our input $\lambda_{h}$ is smaller for the same $M_h$. This will delay the appearance of a plateau and push it to higher scales, allowing more e-folds for lower $M_h$ values. This does not resolve the matter, however, because in both cases the inflationary scale is too high to fit the amplitude of the scalar perturbations. In the slow roll regime this amplitude is given by
\begin{equation}
A_s=\frac{V}{24\pi^2 \epsilon M_{pl}^4} \simeq 2 \times 10^{-9},
\label{eqn: pertubations}
\end{equation}
where for slow roll $\epsilon_{\max}=1$ so we can put and upper bound on the inflationary scale of
\begin{equation}
\max \left(\frac{V^{1/4}}{M_{pl}}\right) \sim 2.5 \times 10^{-2}.
\label{eqn: max_Inflation_scale}
\end{equation}
We find that  whenever enough e-folds are generated by a plateau this upper bound is exceeded.

It is also possible to consider very careful choices of initial conditions such that a large number of e-folds could be generated by the field rolling very slowly passed the inflection point. This was addressed in the slow roll regime in \cite{hamada1308} and it was found that satisfying both the perturbations and the e-folds simultaneously is impossible. Finally, you could imagine repeating the above calculation and allowing for a shallow well to slow the Higgs as it rolls passed, producing more e-folds. It was found that in order to avoid being trapped in the minimum by Hubble friction, the Higgs can only be slowed by a tiny amount. We found that this  impacted the e-folds less than varying $\alpha_s(M_Z)$ by $1 \sigma$.
\begin{Figure}
\centering
\includegraphics[width=\linewidth]{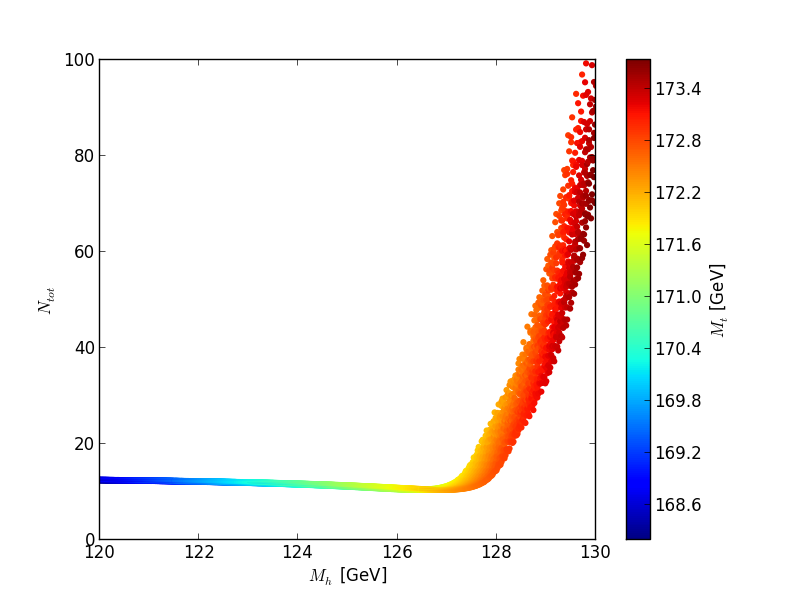}
\captionof{figure}{\it This figure shows the total number of e-folds of inflation caused by a Higgs rolling from rest at $10\ M_{pl}$. The thickness in the band is set by the $\pm 1 \sigma$ error on $\alpha_s(M_Z)$ and the color bar indicates the value of $M_t$ required for a plateau. For smaller $M_h$ the plateau is shorter and occurs at a lower scale and so has only a very small effect on the rolling of the field. For larger $M_h$ the plateau is significant enough to cause an extended period of slow roll.}
\label{fig: plateau_inflation}
\end{Figure}
\section{False Vacuum Inflation}
Although successful inflation cannot be achieved in the simple case of a plateau it may still be possible that the Higgs may be connected to inflation in a slightly less minimal way. To see this we can imagine starting with the plateau situation and increasing the top mass by a few$\times 0.1$ MeV. In this way we can create a false vacuum with large positive energy density that can be used to inflate the universe.

This is then the scenario of old inflation and we are therefore burdened with problem of graceful exit. One possible solution \cite{masina1112} is to extend general relativity to a scalar-tensor theory. This allows the expansion rate of the universe with a constant inflaton energy density to decrease with time, eventually becoming slow enough to allow successful exit through tunnelling. In this paper we revisit an alternative solution, proposed in \cite{masina1204}, that introduces an extra scalar whose dynamics during inflation slowly lifts the Universe out of the false vacuum such that is can roll classically down to the true vacuum. At this point the reader may worry that if we are introducing an extra scalar why we are not just letting that extra scalar to be the inflaton with, say, a quadratic potential. While this is a reasonable position to take, it ignores presence of the false vacuum in the Higgs potential. We also expect that the Higgs will be coupled to any additional scalars (e.g. the scalar responsible for Peccei-Quinn symmetry breaking) that appear above the electroweak scale  through the Higgs portal coupling regardless of whether these scalars can achieve inflation on their own. We therefore consider the approach that the Higgs is responsible for inflation and the additional scalar merely facilitates the graceful exit. For a recent update on false vacuum Higgs inflation see~\cite{Masina1403}, in which the dynamics for the removal of the barrier are left undiscussed.

The tree-level potential in terms of the real-fields is given by,
\begin{equation}
\begin{aligned}
V&=\frac{1}{4} \lambda_s \left(s^2-w^2 \right)^2+\frac{1}{4}\lambda_{h}\left(h^2-v^2\right)^2 \\
&+\frac{1}{4}\lambda_{hs}\left(s^2-w^2 \right)\left(h^2-v^2\right),
\end{aligned}
\end{equation}

where $s$ is the real part of a possibly complex standard model singlet scalar field  and respects a global $\mathds{Z}_2$ (real field) or $U(1)$ (complex field) symmetry. Such a complex field, $S$, arises in the context of invisible axion models, where the symmetry is identified with the $U(1)$ Peccei-Quinn that solves the strong CP problem. The phase of $S$  then becomes the QCD axion.

During inflation the tachyonic $s$ field will roll towards its minimum $\langle s\rangle$ and the mixing term between $h$ and $s$ will grow and lift the false vacuum as shown in figure \ref{fig: end inflation}. The end of inflation is taken as the point at which the false minimum disappears. In reality, tunnelling will become highly probable when the well depth is sufficiently small (when $\Gamma_{\text{tunnel}} \gg H$) causing inflation to end slightly earlier. Additionally the subsequent free rolling of the field down to the global minimum can still produce some inflation. Both of these effects however are small and change the calculation by a negligible number of e-folds which will not alter our conclusions and so we neglect these contributions.
\begin{Figure}
\centering
\includegraphics[width=\linewidth]{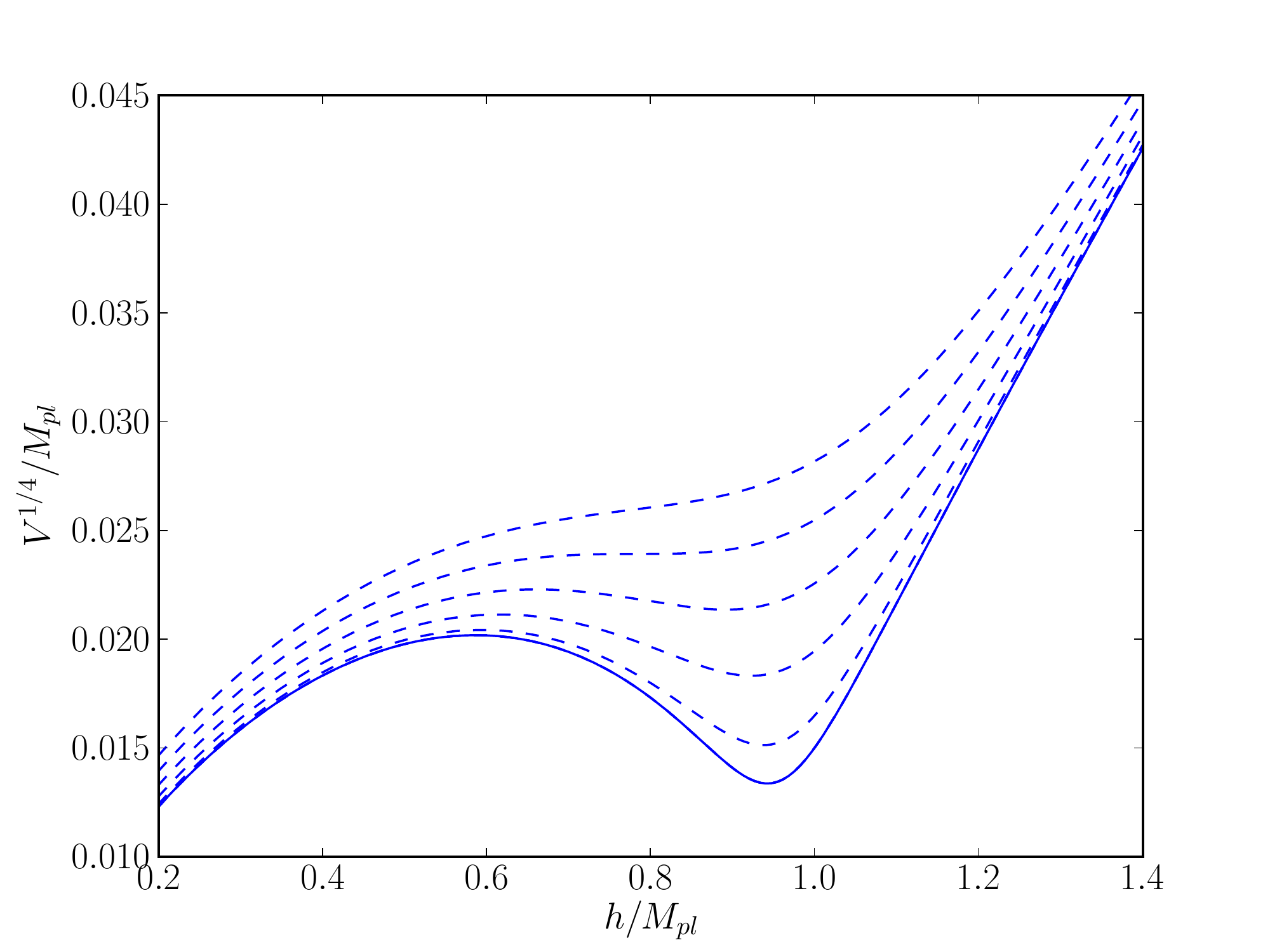}
\captionof{figure}{\it This plot shows the effect of the mixing between the singlet, $s$, and the Higgs, $h$, on the Higgs contribution to $V$ as $s$ rolls towards its minimum. The singlet field manages to successfully remove the false vacuum allowing the Higgs to roll down to true vacuum}
\label{fig: end inflation}
\end{Figure}
We therefore have a setup similar to hybrid potential \cite{Linde9307} in which the rolling of $s$ triggers the waterfall field $h$ but in this case the false vacuum is created purely by quantum effects. It is possible to treat this as an approximately single field model because $h$ is trapped in the false minimum throughout inflation. The renormalized potential can then be written\footnote{Here we neglect the $h^2$ term because we are interested in the large $h$ behaviour.} as a function of $s$
\begin{equation}
\begin{aligned}
V_s&=\frac{1}{4}\lambda_s \left( s^2-\langle s \rangle^2\right)^2\\&+\frac{1}{4}\left( \lambda_{\text{eff}} -\frac{\lambda_{hs}^2}{4 \lambda_s}\right) \left(\langle h \rangle^2 -v^2\right)^2,
\end{aligned}
\end{equation}
with
\begin{equation}
\langle s \rangle =\frac{1}{\sqrt{2 \lambda_s}}\sqrt{M_s^2+\lambda_{hs}(v^2-\langle h \rangle^2)}\ ,
\end{equation}
where $M_s$ is the mass of the new scalar and $v=246$ GeV is Higgs vev in the true vacuum. The position, $\langle h \rangle$, of the false vacuum will change during inflation (see figure \ref{fig: end inflation}) so we may treat it as a function of $s$.

As $h$ rolls to the global minimum $\langle s \rangle$ relaxes to its ground state value given by
\begin{equation}
f_a =\left. \langle s \rangle \right|_{\langle h \rangle =v}=\frac{M_s}{\sqrt{2 \lambda_s}},
\end{equation}
where $f_a$ can be interpreted as the axion decay constant for the case where $S$ is a complex field charge under $U(1)_{PQ}$.

The total number of e-folds is then calculated using 
\begin{equation}
N=\frac{1}{M_{pl}} \int_{s=0}^{s_{end}} \frac{ds}{\sqrt{2 \epsilon}}\ ,
\end{equation}
with 
\begin{equation}
\epsilon=\frac{M_{pl}^2}{2} \left( \frac{V'_s}{V_s}\right)^2 .
\end{equation}
The amplitude of the scalar perturbations are then calculated $N_*=50-60$ e-folds before the end of inflation using equation (\ref{eqn: pertubations}).

The introduction of the new scalar will modify the low energy Higgs parameters and the RGE's of standard model. Firstly, it was shown in \cite{eliasMiro1203} that when the mass of the extra scalar is much larger than the electroweak scale (as will always be the case here) it can be integrated out to yield an effective theory below $M_s$. In this effective theory the Higgs quartic coupling will be modified to that of the standard model as a result of the $\lambda_{hs}$ mixing term. Below $M_s$ we must replace 
\begin{equation}
\lambda_h \rightarrow \lambda=\lambda_h -\frac{\lambda_{hs}^2}{4 \lambda_s},
\end{equation}
where $\lambda \sim 0.129$ is what is inferred from the Higgs mass measurement and is what enters the SM running below $M_s$. At $M_s$ we must therefore apply a threshold effect to match to the full theory by replacing $\lambda$ with $\lambda_h=\lambda +\frac{\lambda_{hs}^2}{4 \lambda_s}$. Above $M_s$ we must also include the $s$ field in the RGE's
\begin{align}
(4\pi)^2&\beta_{\lambda_{h}} = (4\pi)^2\beta_{\lambda_{h}}^{SM}+\frac{1}{2}\lambda_{hs}^2,\\
(4\pi)^2&\beta_{\lambda_{hs}} = \frac{1}{4}\lambda_{hs}\left(12y_t^2-\frac{9}{5}g_1^2-9g_2^2\right)\\ \nonumber
& \hspace{1cm}+\lambda_{hs}(6\lambda_h+4\lambda_s)+2\lambda_{hs}^2,\\
(4\pi)^2&\beta_{\lambda_{s}} = \lambda_{hs}^2+10\lambda_s^2.
\end{align}
In order for this mechanism to work both $\lambda_s$ and $\lambda_{hs}$ will need to be very small so the RGE contribution will be minor. The threshold effect however can still be significant because even small changes in $\lambda_h$ can substantially change the position of the false vacuum.

In order to test this model we select as inputs \{$M_h,\alpha_s(M_Z), \lambda_s,\lambda_{hs}, M_s$\}. Requiring the presence of the false vacuum then largely determines $M_t$. There is, however, a degeneracy in such an approach resulting from the freedom of tuning the initial depth of the well using $M_t$. Different well depths will result in different $s_{end}$, and hence $A_s$, values for the same set of input parameters. To resolve this we tune $M_t$  for each set of inputs such that the resulting well depth generates the best possible fit to  $A_s$. We therefore choose only the best possible point in the degenerate set of outputs given our 5 inputs.

The result of an extensive nested sampling scan using MultiNest \cite{Feroz0809} is shown in figure \ref{fig: as_mh_res}. The $\chi^2$ was derived from fitting the experimental value of $M_h=125.66 \pm 0.34$ GeV (see \cite{degrassi1205} and references therein) and $M_t=173.35 \pm 0.76$ GeV \cite{ATLAS1403}, the world average of $\alpha_s(M_Z)=0.1184 \pm 0.0007$ \cite{Bethke0908}, and observed value of $A_s=(2.196 \pm 0.060) \times 10^{-9}$ \cite{Planck}. It is clear from figure \ref{fig: as_mh_res} that the best fit point is inconsistent with the 3 sigma contours in $(M_h,\alpha_s(M_Z))$. 

\begin{Figure}
\centering
\includegraphics[width=\linewidth]{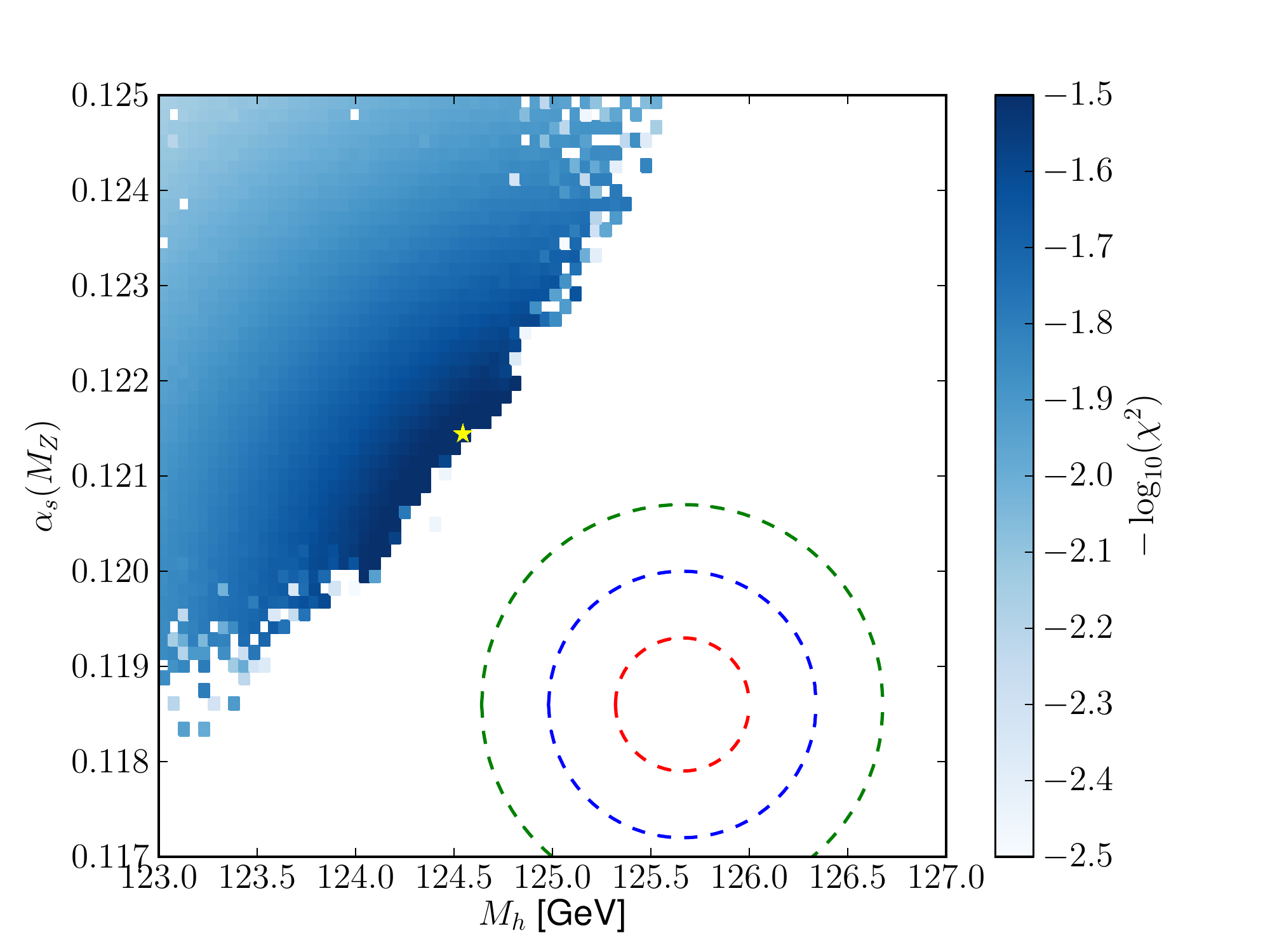}
\captionof{figure}{\it This plot shows the binned best fit point in the $M_h$-$\alpha_s(M_Z)$ plane for $N_*=50$.  Also depicted are the $1$,$2$, and $3 \sigma$ experimental contours. The global best fit point, marked with a yellow star, is inconsistent with experiment at more that $3 \sigma$. }
\label{fig: as_mh_res}
\end{Figure}

There is a clearly visible sharp boundary in figure \ref{fig: as_mh_res}. This marks the point when the scale of inflation exceeds the maximum value allowed by equation (\ref{eqn: max_Inflation_scale}) preventing any chance of achieving a good fit to $A_s$ by tweaking $\epsilon$. The corresponding distributions for $\lambda_{s}, \lambda_{hs}$, and $M_s$ and their best fit value are shown in figure \ref{fig: params}. It is clear that in order to fit the scalar perturbations and the e-folds simultaneously, quite small values for the couplings are needed, which would result in a very large (Planckian) expectation value $\left. \langle s \rangle \right|_{\langle h \rangle =v}$ for the extra scalar field, making the case where we identify it with the real part of the Peccei-Quinn scale less phenomenologically interesting. The sharp line in the $\lambda_{s}$-$\lambda_{hs}$ plane marks the region above which the threshold effect becomes too large and results in a push of the false vacuum to too large scales.

\begin{Figure}
\centering
\includegraphics[width=1.\linewidth]{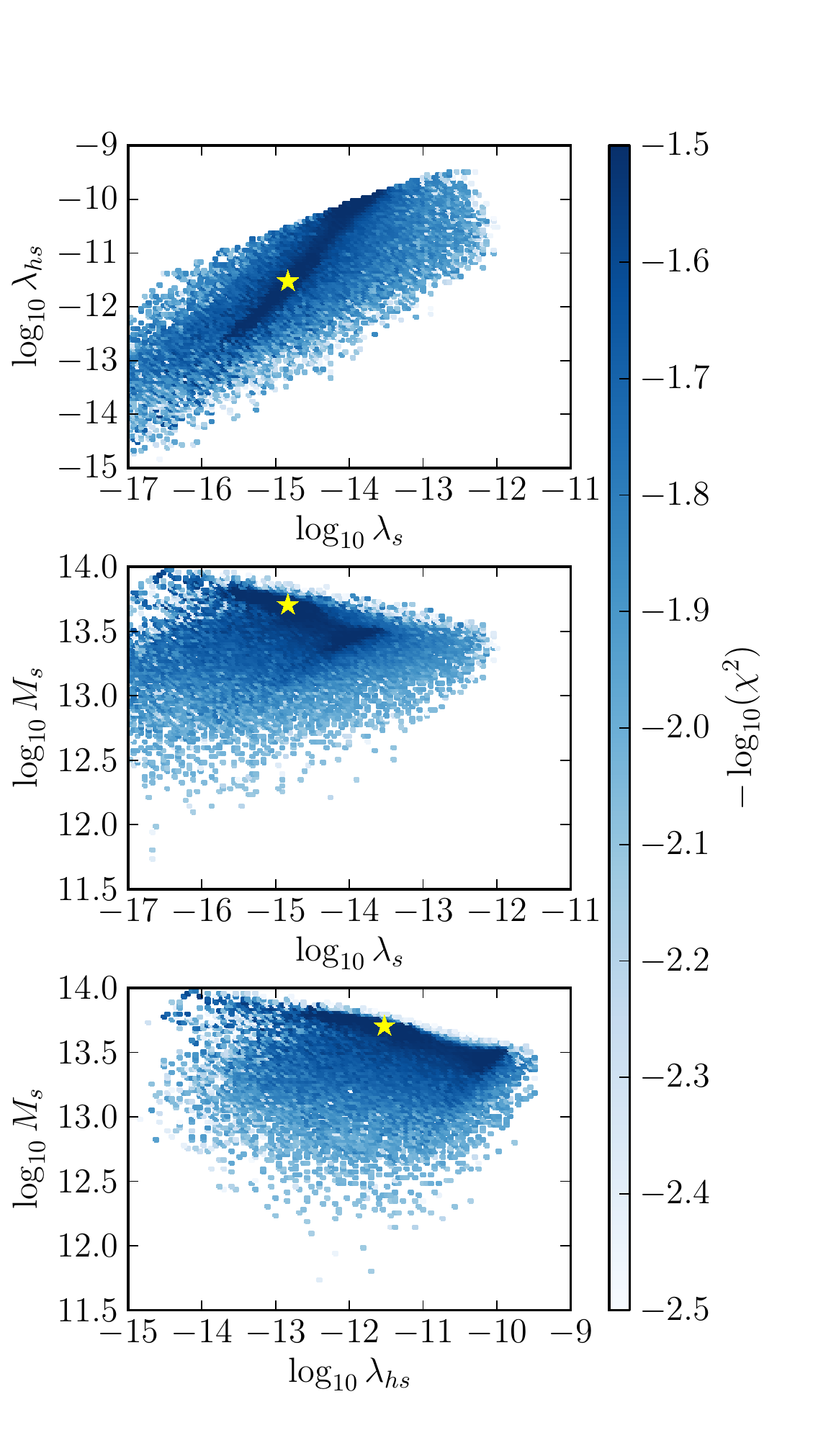}
\captionof{figure}{\it This figure shows the binned best fit point for $\lambda_{s}, \lambda_{hs}$, and $M_s$ with $N_*=50$.  The global best fit point is marked with a yellow star.}
\label{fig: params}
\end{Figure}

\section{Conclusion}
In this paper we have considered two possible implementations of minimal Higgs inflation. In section 2 we tuned the Higgs potential in such a way that a plateau appears and investigated whether this plateau can be used to inflate the universe via a slow-rolling of the Higgs alone. We considered the full 3-loop RGE improved 2-loop effective potential. A simultaneous fit of the number of e-foldings and the scalar perturbations turned out to be impossible, such that an extension of the Standard Model is necessary, compare figure 2.

The most minimal extension was investigated in section 3 where we introduced an additional singlet scalar field $s$ and looked at a hybrid scenario. Such a scalar field is motivated by the strong CP-problem. In this case, the Higgs sits in a local minimum of the potential and $s$ slowly rolls towards the minimum of its potential. The mutual coupling between $s$ and the Higgs field removes the barrier during the rolling of $s$ such that the Higgs can then roll towards its global minimum and successful exit is guaranteed. To ensure a correct treatment, we included the 1-loop RGE's for the new scalar, the threshold effect in the Higgs potential occurring at the mass of the singlet scalar, the movement of the Higgs field during inflation and the degeneracy in the well depth.

Our results are summarised in figure 4 where one can see that those sets of parameters that give a good fit to the inflationary observables are clearly excluded by measurements of the Higgs mass and the strong coupling constant. 

With standard General Relativity and Quantum Field Theory with minimal couplings between the particles and gravity it has been shown that one cannot obtain inflation using only the standard model Higgs.  In this work we show that even with an additional field allowing the Higgs to become the waterfall field of a hybrid inflation model, the coupling between the two fields conspires to prevent good inflationary parameters.  Inflation can only be explained using either a more complicated scenario or an entirely separate field such that the Higgs plays no role in the process.

\section*{Acknowledgements}
MF is funded by the STFC, PG is funded on an ERC Senior Fellowship and RH receives funding from a KCL graduate school award.

\bibliographystyle{h-physrev}
\bibliography{bibliography}

\begin{thebibliography}{10}

\bibitem{Planck}
Planck Collaboration, P.~Ade {\em et~al.},
\newblock (2013), 1303.5082.

\bibitem{BICEP}
BICEP2 Collaboration, P.~Ade {\em et~al.},
\newblock (2014), 1403.3985.

\bibitem{ATLAS1207}
ATLAS Collaboration, G.~Aad {\em et~al.},
\newblock Phys.Lett. {\bf B716}, 1 (2012), 1207.7214.

\bibitem{CMS1207}
CMS Collaboration, S.~Chatrchyan {\em et~al.},
\newblock Phys.Lett. {\bf B716}, 30 (2012), 1207.7235.

\bibitem{Ellis1303}
J.~Ellis and T.~You,
\newblock JHEP {\bf 1306}, 103 (2013), 1303.3879.

\bibitem{Isidori0712}
G.~Isidori, V.~S. Rychkov, A.~Strumia, and N.~Tetradis,
\newblock Phys.Rev. {\bf D77}, 025034 (2008), 0712.0242.

\bibitem{Bezrukov0710}
F.~L. Bezrukov and M.~Shaposhnikov,
\newblock Phys.Lett. {\bf B659}, 703 (2008), 0710.3755.

\bibitem{Bezrukov0812}
F.~L. Bezrukov, A.~Magnin, and M.~Shaposhnikov,
\newblock Phys.Lett. {\bf B675}, 88 (2009), 0812.4950.

\bibitem{Bezrukov0812a}
F.~Bezrukov, D.~Gorbunov, and M.~Shaposhnikov,
\newblock JCAP {\bf 0906}, 029 (2009), 0812.3622.

\bibitem{Barbon0903}
J.~Barbon and J.~Espinosa,
\newblock Phys.Rev. {\bf D79}, 081302 (2009), 0903.0355.

\bibitem{Bezrukov0904}
F.~Bezrukov and M.~Shaposhnikov,
\newblock JHEP {\bf 0907}, 089 (2009), 0904.1537.

\bibitem{Lerner0912}
R.~N. Lerner and J.~McDonald,
\newblock JCAP {\bf 1004}, 015 (2010), 0912.5463.

\bibitem{Burgess1002}
C.~Burgess, H.~M. Lee, and M.~Trott,
\newblock JHEP {\bf 1007}, 007 (2010), 1002.2730.

\bibitem{Bezrukov1008}
F.~Bezrukov, A.~Magnin, M.~Shaposhnikov, and S.~Sibiryakov,
\newblock JHEP {\bf 1101}, 016 (2011), 1008.5157.

\bibitem{Giudice1010}
G.~F. Giudice and H.~M. Lee,
\newblock Phys.Lett. {\bf B694}, 294 (2011), 1010.1417.

\bibitem{burgess1402}
C.~Burgess, S.~P. Patil, and M.~Trott,
\newblock (2014), 1402.1476.

\bibitem{Cook1403}
J.~L. Cook, L.~M. Krauss, A.~J. Long, and S.~Sabharwal,
\newblock (2014), 1403.4971.

\bibitem{Bezrukov1403}
F.~Bezrukov and M.~Shaposhnikov,
\newblock (2014), 1403.6078.

\bibitem{Hamada1403}
Y.~Hamada, H.~Kawai, K.-y. Oda, and S.~C. Park,
\newblock (2014), 1403.5043.

\bibitem{masina1204}
I.~Masina and A.~Notari,
\newblock JCAP {\bf 1211}, 031 (2012), 1204.4155.

\bibitem{degrassi1205}
G.~Degrassi {\em et~al.},
\newblock JHEP {\bf 1208}, 098 (2012), 1205.6497.

\bibitem{buttazzo1307}
D.~Buttazzo {\em et~al.},
\newblock JHEP {\bf 1312}, 089 (2013), 1307.3536.

\bibitem{Isidori0104}
G.~Isidori, G.~Ridolfi, and A.~Strumia,
\newblock Nucl.Phys. {\bf B609}, 387 (2001), hep-ph/0104016.

\bibitem{ArkaniHamed0801}
N.~Arkani-Hamed, S.~Dubovsky, L.~Senatore, and G.~Villadoro,
\newblock JHEP {\bf 0803}, 075 (2008), 0801.2399.

\bibitem{Ellis0906}
J.~Ellis, J.~Espinosa, G.~Giudice, A.~Hoecker, and A.~Riotto,
\newblock Phys.Lett. {\bf B679}, 369 (2009), 0906.0954.

\bibitem{EliasMiro1112}
J.~Elias-Miro {\em et~al.},
\newblock Phys.Lett. {\bf B709}, 222 (2012), 1112.3022.

\bibitem{Espinosa0710}
J.~Espinosa, G.~Giudice, and A.~Riotto,
\newblock JCAP {\bf 0805}, 002 (2008), 0710.2484.

\bibitem{Lebedev1210}
O.~Lebedev and A.~Westphal,
\newblock Phys.Lett. {\bf B719}, 415 (2013), 1210.6987.

\bibitem{Kobakhidze1301}
A.~Kobakhidze and A.~Spencer-Smith,
\newblock Phys.Lett. {\bf B722}, 130 (2013), 1301.2846.

\bibitem{Fairbairn1403}
M.~Fairbairn and R.~Hogan,
\newblock (2014), 1403.6786.

\bibitem{hamada1308}
Y.~Hamada, H.~Kawai, and K.-y. Oda,
\newblock PTEP {\bf 2014}, 023B02 (2014), 1308.6651.

\bibitem{masina1112}
I.~Masina and A.~Notari,
\newblock Phys.Rev. {\bf D85}, 123506 (2012), 1112.2659.

\bibitem{Masina1403}
I.~Masina,
\newblock (2014), 1403.5244.

\bibitem{Linde9307}
A.~D. Linde,
\newblock Phys.Rev. {\bf D49}, 748 (1994), astro-ph/9307002.

\bibitem{eliasMiro1203}
J.~Elias-Miro, J.~R. Espinosa, G.~F. Giudice, H.~M. Lee, and A.~Strumia,
\newblock JHEP {\bf 1206}, 031 (2012), 1203.0237.

\bibitem{Feroz0809}
F.~Feroz, M.~Hobson, and M.~Bridges,
\newblock Mon.Not.Roy.Astron.Soc. {\bf 398}, 1601 (2009), 0809.3437.

\bibitem{ATLAS1403}
ATLAS Collaboration, CDF Collaboration, CMS Collaboration, D0 Collaboration,
\newblock (2014), 1403.4427.

\bibitem{Bethke0908}
S.~Bethke,
\newblock Eur.Phys.J. {\bf C64}, 689 (2009), 0908.1135.

\end{thebibliography}
\end{multicols}
\end{document}